\documentclass{PoS}

\usepackage{graphicx}
\usepackage{subfig}
\usepackage{float}
\usepackage{amsmath}
\usepackage{amssymb}
\usepackage{fontenc}
\usepackage{times}
\usepackage{mathptmx}
\usepackage{amsthm}
\usepackage{latexsym}
\usepackage{dcolumn}
\usepackage{color}

\newcommand{\newc}{\newcommand*}

\long\def\begincomment#1\endcomment{%
        \begingroup\sf\baselineskip12pt#1\endgroup}

\newc{\etal}{\textrm{et al.}} 
\newc{\eg}{\textrm{e.g.}} 
\newc{\ie}{\textrm{i.e.}}
\newc{\etc}{\textrm{etc}}
\newc\vs{\textrm{vs.}}
\newc{\cl}{\rm {CL}}

\newc{\ev}{\ensuremath{\,\mathrm{eV}}}
\newc{\kev}{\ensuremath{\,\mathrm{keV}}}
\newc{\mev}{\ensuremath{\,\mathrm{MeV}}}
\newc{\gev}{\ensuremath{\,\mathrm{GeV}}}
\newc{\tev}{\ensuremath{\,\mathrm{TeV}}}

\newc{\MeV}{\mev} 
\newc{\TeV}{\tev}
\newc{\invpb}{\ensuremath{/\text{pb}}}
\newc{\invfb}{\ensuremath{/\text{fb}}}

\newc\nb{\ensuremath{\,\mathrm{nb}}} \newc\pb{\ensuremath{\,\mathrm{pb}}} \newc\fb{\ensuremath{\,\mathrm{fb}}}

\newc\pc{\ensuremath{\,\mathrm{pc}}}
\newc\kpc{\ensuremath{\,\mathrm{kpc}}}
\newc\mpc{\ensuremath{\,\mathrm{Mpc}}}

\newc\ps{\ensuremath{\,\mathrm{ps}}} 

\newc\cmeter{\ensuremath{\,\mathrm{cm}}} 
\newc\meter{\ensuremath{\,\mathrm{m}}} 
\newc\kmeter{\ensuremath{\,\mathrm{km}}}

\newc\second{\ensuremath{\,\mathrm{s}}}
\newc\msecond{\ensuremath{\,\mathrm{ms}}}
\newc\nsecond{\ensuremath{\,\mathrm{ns}}}
\newc\psecond{\ensuremath{\,\mathrm{ps}}}

\newc{\chisqmin}{\ensuremath{\chi^2_{\mathrm{min}}}}
\newc{\Delchisq}{\ensuremath{\Delta\chi^2}}
\newc{\chisq}{\ensuremath{\chi^2}}

\newc{\like}{\ensuremath{\mathcal{L}}}

\newc\lsim{\ensuremath{\mathrel{\rlap{\lower4pt\hbox{\hskip1pt$\sim$}}\raise1pt\hbox{$<$}}}}
\newc\gsim{\ensuremath{\mathrel{\rlap{\lower4pt\hbox{\hskip1pt$\sim$}}\raise1pt\hbox{$>$}}}}

\newc{\VEV}[1]{\ensuremath{\langle #1 \rangle}}

\newc{\dl}{\ensuremath{\stackrel{\leftarrow}{D}}}
\newc{\dr}{\ensuremath{\stackrel{\rightarrow}{D}}}

\newc{\bcenter}{\begin{center}}    \newc{\ecenter}{\end{center}}
\newc{\bfl}{\begin{flushleft}}    \newc{\efl}{\end{flushleft}}
\newc{\bfr}{\begin{flushright}}    \newc{\efr}{\end{flushright}}

\newc{\bi}{\begin{itemize}}
\newc{\ei}{\end{itemize}}
\newc{\bed}{\begin{description}}
\newc{\eed}{\end{description}}
\newc{\ben}{\begin{enumerate}}
\newc{\een}{\end{enumerate}}

\newc{\be}{\begin{equation}}
\newc{\ee}{\end{equation}}
\newc{\bea}{\begin{eqnarray}}
\newc{\eea}{\end{eqnarray}}
\newc{\ra}{\rightarrow}

\newc{\alphas}{\ensuremath{\alpha_s}}
\newc{\alphatwo}{\ensuremath{\alpha_2}}
\newc{\alphaone}{\ensuremath{\alpha_1}}
\newc{\alphai}[1]{\ensuremath{\alpha_{#1}}}
\newc{\alphaem}{\ensuremath{\alpha_{\mathrm{em}}}}

\newc{\alphaeff}{\ensuremath{\alpha_{\mathrm{eff}}}}

\newc{\sineff}{\ensuremath{\sin \theta_{\mathrm{eff}}}}
\newc{\sinsqeff}{\ensuremath{\sin^2 \theta_{\mathrm{eff}}}}
\newc{\dalphahad}{\ensuremath{\Delta \alpha_{\mathrm{had}}}}

\newc{\yt}{\ensuremath{h_t}} \newc{\yb}{\ensuremath{h_b}} \newc{\ytau}{\ensuremath{h_{\tau}}}

\newc\mz{\ensuremath{M_Z}} 
\newc\mw{\ensuremath{m_W}}
\newc\mZ{\mz}        \newc\mW{\mw}

\newc\mhsm{\ensuremath{ m_{H_{\mathrm{SM}}}}}

\newc{\mtop}{\ensuremath{ m_t}}               \newc{\mtpole}{\ensuremath{ M_t}}
\newc{\mbottom}{\ensuremath{ m_b}} 
\newc{\mtau}{\ensuremath{ m_{\tau}}}
\newc{\mt}{\mtpole}
\newc{\mb}{\mbottom} 

\newc{\rtwogg}{\ensuremath{R_{h_2}(\gamma\gamma)}}
\newc{\rtwozz}{\ensuremath{R_{h_2}(ZZ)}}
\newc{\ronegg}{\ensuremath{R_{h_1}(\gamma\gamma)}}
\newc{\ronezz}{\ensuremath{R_{h_1}(ZZ)}}
\newc{\rsiggg}{\ensuremath{R_{h_\textrm{sig}}(\gamma\gamma)}}
\newc{\rsigzz}{\ensuremath{R_{h_\textrm{sig}}(ZZ)}}

\newc{\llbar}{\ensuremath{\ell\bar{\ell}}}
\newc{\tauptaum}{\ensuremath{ \tau^+\tau^-}}

\newc{\qqbar}{\ensuremath{ q\bar{q}}} \newc{\ppbar}{\ensuremath{ p\bar{p}}}
\newc{\bbbar}{\ensuremath{ b\bar{b}}} \newc{\ttbar}{\ensuremath{ t\bar{t}}}
\newc{\ffbar}{\ensuremath{ f\bar{f}}} \newc{\tautaubar}{\ensuremath{ \tau\bar{\tau}}}

\newc{\mchi}{\ensuremath{m_\neutone}}
\newc{\squark}{\ensuremath{\tilde{q}}}
\newc{\slepton}{\ensuremath{\tilde{l}}}
\newc{\gluino}{\ensuremath{\tilde{g}}} 
\newc{\mgluino}{\ensuremath{{m_{\gluino}}}}

\newc{\sthw}{\ensuremath{ \sin\theta_W}}              \newc{\cthw}{\ensuremath{\cos\theta_W}}
\newc{\tanthw}{\ensuremath{ \tan\theta_W}}              \newc{\cotthw}{\ensuremath{\cot\theta_W}}

\newc{\ssqthw}{\ensuremath{\sin^2 \theta_W}}

\newc{\msbar}{\ensuremath{\overline{MS}}} \newc{\drbar}{\ensuremath{\overline{DR}}}

\newc{\mtmtsmmsbar}{\ensuremath{ m_t(m_t)^{\msbar}_{{\mathrm{SM}}}}}
\newc{\mtmtsmdrbar}{\ensuremath{ m_t(m_t)^{\drbar}_{{\mathrm{SM}}}}}
\newc{\mtmtmssmdrbar}{\ensuremath{ m_t(m_t)^{\drbar}_{{\mathrm{SUSY}}}}}

\newc{\mbmbmsbar}{\ensuremath{ m_b(m_b)^{\msbar} }}

\newc{\mbmbsmmsbar}{\ensuremath{ m_b(m_b)^{\msbar}_{{\mathrm{SM}}}}}
\newc{\mbmzsmmsbar}{\ensuremath{ m_b(\mz)^{\msbar}_{{\mathrm{SM}}}}}
\newc{\mbmzsmdrbar}{\ensuremath{ m_b(\mz)^{\drbar}_{{\mathrm{SM}}}}}
\newc{\mbmzmssmdrbar}{\ensuremath{ m_b(\mz)^{\drbar}_{{\mathrm{SUSY}}}}}

\newc{\mtaumzsmmsbar}{\ensuremath{ m_{\tau}(\mz)^{\msbar}_{{\mathrm{SM}}}}}
\newc{\mtaumzsmdrbar}{\ensuremath{ m_{\tau}(\mz)^{\drbar}_{{\mathrm{SM}}}}}
\newc{\mtaumzmssmdrbar}{\ensuremath{ m_{\tau}(\mz)^{\drbar}_{{\mathrm{SUSY}}}}}

\newc{\alphasmzms}{\ensuremath{\alpha_s(M_Z)^{\overline{MS}}}}
\newc{\alphaimzms}[1]{\ensuremath{\alpha_{#1}(M_Z)^{\overline{MS}}}}

\newc{\alphaemmz}{\ensuremath{\alpha_{\mathrm{em}}(M_Z)^{\overline{MS}}}}

\newc{\mzero}{\ensuremath{{m_0}}}
\newc{\mhalf}{\ensuremath{ m_{1/2}}}
\newc{\tanb}{\ensuremath{\tan\beta}}
\newc{\azero}{\ensuremath{ A_0}}
\newc{\signmu}{\ensuremath{\rm{sgn}\,\mu}}
\newc{\atau}{\ensuremath{{A_{\tau}}}}

\newc{\mueff}{\ensuremath{\mu_{\rm{eff}}}}

\newc{\lam}{\ensuremath{{\lambda}}}
\newc{\kap}{\ensuremath{{\kappa}}}

\newc{\alam}{\ensuremath{{A_{\lambda}}}}
\newc{\akap}{\ensuremath{{A_{\kappa}}}}

 \newc{\hs}{\ensuremath{ H_s}}      
\newc{\mhs}{\ensuremath{ m_{H_s}}} 

\newc{\mgut}{\ensuremath{ M_{\rm GUT}}}
\newc{\mplanck}{\ensuremath{ M_{\rm P}}}      \newc{\mpl}{\ensuremath{ M_{\rm Pl}}}
\newc{\msusy}{\ensuremath{ M_{\rm SUSY}}}      \newc{\ms}{\ensuremath{ M_{\rm S}}}

 \newc{\hu}{\ensuremath{ H_u}}       \newc{\hd}{\ensuremath{ H_d}}
 \newc{\mhu}{\ensuremath{ m_{H_u}}}       \newc{\mhd}{\ensuremath{ m_{H_d}}}
 \newc{\mhuew}{\ensuremath{ m^{\ast}_{H_u}}}       \newc{\mhdew}{\ensuremath{ m^{\ast}_{H_d}}}
 \newc{\mhuewsq}{\ensuremath{ m^{\ast\, 2}_{H_u}}}       \newc{\mhdewsq}{\ensuremath{ m^{\ast\, 2}_{H_d}}}
 \newc{\mhl}{\ensuremath{m_\hl}} 
 \newc{\mhone}{\ensuremath{m_{h_1}}} 
 \newc{\mhtwo}{\ensuremath{m_{h_2}}} 
 \newc{\mglu}{\ensuremath{m_{\tilde g}}} 
 \newc{\mul}{\ensuremath{m_{\tilde{u}_L}}} 
 \newc{\mtone}{\ensuremath{m_{\tilde{t}_1}}} 
 \newc{\ma}{\ensuremath{m_A}} 
 \newc{\maone}{\ensuremath{m_{a_1}}} 
 \newc{\matwo}{\ensuremath{m_{a_2}}}
 \newc{\hone}{\ensuremath{h_1}}
 \newc{\htwo}{\ensuremath{h_2}}
 \newc{\aone}{\ensuremath{a_1}}
 \newc{\atwo}{\ensuremath{a_2}}

\newc{\sigsip}{\ensuremath{\sigma^{\rm SI}_{p}}}	\newc{\sigsin}{\ensuremath{\sigma^{\rm SI}_{n}}}
\newc{\sigsdp}{\ensuremath{\sigma^{\rm SD}_{p}}}	\newc{\sigsdn}{\ensuremath{\sigma^{\rm SD}_{n}}}
\newc{\sigsi}{\ensuremath{\sigma^{\rm SI}}}	\newc{\sigsd}{\ensuremath{\sigma^{\rm SD}}}

\newc{\abund}{\ensuremath{ \Omega h^2}}
\newc{\omegadm}{\ensuremath{ \Omega_{{\rm DM}}}}     \newc{\abunddm}{\ensuremath{ \Omega_{{\rm DM}} h^2}} 
\newc{\omegam}{\ensuremath{ \Omega_{{\rm m}}}}       \newc{\abundm}{\ensuremath{ \Omega_{{\rm m}} h^2}}
\newc{\omegab}{\ensuremath{ \Omega_{{\rm b}}}}	\newc{\abundb}{\ensuremath{ \Omega_{{\rm b}} h^2}}
\newc{\omegatot}{\ensuremath{ \Omega_{{\rm TOT}}}}
\newc{\omegacdm}{\ensuremath{ \Omega_{{\rm CDM}}}}   \newc{\abundcdm}{\ensuremath{ \Omega_{{\rm CDM}} h^2}}
\newc{\omegalambda}{\ensuremath{ \Omega_{\Lambda}}} \newc{\abundlambda}{\ensuremath{ \Omega_{\Lambda} h^2}}
\newc{\omegarad}{\ensuremath{ \Omega_{{\rm rad}}}}  \newc{\abundrad}{\ensuremath{ \Omega_{{\rm rad}} h^2}}

\newc{\rhocrit}{\ensuremath{ \rho_{\rm crit}}}
\newc{\rhochi}{\ensuremath{ \rho_{\chi}}}

\newc{\abunchi}{\ensuremath{\Omega_\chi h^2}}
\newc{\abundlsp}{\ensuremath{\Omega_{\rm LSP}h^2}}

\newc{\amu}{\ensuremath{ a_{\mu}}}        \newc{\amususy}{\ensuremath{ a_{\mu}^{\mathrm{SUSY}}}}
\newc{\amuexpt}{\ensuremath{ a_{\mu}^{\mathrm{expt}}}}        \newc{\amusm}{\ensuremath{ a_{\mu}^{\mathrm{SM}}}}
\newc\deltaamu{\ensuremath{\Delta a_{\mu}}} \newc{\deltaamususy}{\ensuremath{\delta a_{\mu}^{\mathrm{SUSY}}}}
\newc\gmtwo{\ensuremath{ (g-2)_{\mu}}} 
\newc{\deltagmtwomususy}{\ensuremath{\delta\left(g-2\right)_{\mu}^{\mathrm{SUSY}}}}
\newc{\deltagmtwomu}{\ensuremath{\delta\left(g-2\right)_{\mu}}}

\newc\BR{\ensuremath{\rm BR}}

\newc\bsgamma{\ensuremath{ b\rightarrow s \gamma }}
\newc\bxsgamma{\ensuremath{\overline{B}\rightarrow X_{s}\gamma}}

\newc\brbsgamma{\ensuremath{\BR\left(\bsgamma\right)}}
\newc\brbxsgamma{\ensuremath{\BR\left(\bxsgamma\right)}}

\newc\bsmumu{\ensuremath{B_s\to\mu^+\mu^-}}
\newc\brbsmumu{\ensuremath{\BR\left(B_s\to\mu^+\mu^-\right)}}

\newc\bdmmumu{\ensuremath{\overline{B}_d\to\mu^+\mu^-}}

\newc\bbbarmix{\ensuremath{\overline{B}_s\mbox{-}B_s}}      
\newc\delmbs{\ensuremath{\Delta M_{B_s}}}

\newc{\butaunu}{\ensuremath{B_u \rightarrow \tau \nu}}
\newc{\brbutaunu}{\ensuremath{\BR\left(B_u \rightarrow \tau \nu\right)}}

\newcommand*{\neutone}{\ensuremath{\chi}}

\let\oldcite\cite
\renewcommand*{\cite}{~\oldcite}

\newcommand*{\hl}{\ensuremath{h}}

\title{Revisiting a light NMSSM pseudoscalar at the LHC}

\ShortTitle{Revisiting a light NMSSM pseudoscalar at the LHC}

\author{Nils-Erik Bomark,$^a$ Stefano Moretti,$^b$
    \speaker{Shoaib Munir}$^c$ and Leszek Roszkowski$^a$ \\
\llap{$^a$}National Centre for Nuclear Research\\
  Ho{\. z}a 69, 00-681 Warsaw, Poland \\     
\llap{$^b$}School of Physics \& Astronomy, University of Southampton \\
Southampton SO17 1BJ, UK \\   
\llap{$^c$}Department of Physics and Astronomy, Uppsala University \\
Box 516, SE-751 20 Uppsala, Sweden \\
        E-mail: \email{nbomark@fuw.edu.pl}, \email{s.moretti@soton.ac.uk}, \email{shoaib.munir@physics.uu.se}, \email{leszek.roszkowski@fuw.edu.pl}}

\abstract{The discovery of a light, singlet-like pseudoscalar Higgs
  boson, $A_1$, of the Next-to-Minimal Supersymmetric Standard Model (NMSSM)
    could provide a hallmark signature of non-minimal supersymmetry. 
We review here the potential of
the LHC to probe such a light $A_1$ in the decays of
one of the heavier scalar Higgs bosons of the NMSSM. 
We find the production of pairs of the $A_1$, with a mass below
 60\gev\ or so, via decays of the two lightest scalar states to be
 especially promising, for an integrated luminosity as low as
30/fb. For heavier masses, the decay of the
heaviest scalar into a $Z$ boson and an $A_1$ could 
lead to its detection at the LHC.}

\FullConference{Prospects for Charged Higgs Discovery at Colliders - CHARGED 2014,\\
		16-18 September 2014\\
		Uppsala University, Sweden}

\begin{document}

\section{\label{intro}Introduction}

The NMSSM contains an extra singlet Higgs
superfield in addition to the two doublet superfields of the Minimal
Supersymmetric Standard Model. As a result, there are a total of five
neutral Higgs mass eigenstates: scalars $H_i$, with $i=1,2,3$, and pseudoscalars
$A_{1,2}$, and a charged pair $H^\pm$, in the model.
The masses of the two new singlet-like states are generally very weakly
constrained by the Higgs boson data from the Large Electron Positron
 collider or the Large Hadron Collider (LHC), and can be
as low as a few \gev. We assess the scope of the detectability of a
light, $\lesssim 150$\gev, $A_1$ of the NMSSM at the run 2 of the LHC
 with $\sqrt{s}=14$\tev. Through dedicated scans of the 
parameter space of the contrained NMSSM with
non-universal Higgs masses (CNMSSM-NUHM), we found a considerable
number of points containing such $A_1$ while also satisfying
important experimental constraints. We then performed a 
detailed signal-to-background analysis for each of the main production 
and decay channels of $A_1$. Most notably, we employed the jet substructure
method for detecting the $b$-quarks originating from $A_1$ decays,
which considerably improves the experimental sensitvitiy.
 
\section{\label{model} $A_1$ production channels in the model studied}

The soft supersymmetry (SUSY)-breaking Higgs potential of the NMSSM is written as
\begin{eqnarray}
V_{\rm{soft}}=m_{\hu}^2|\hu|^2+m_{\hd}^2|\hd|^2+m_{S}^2|S|^2+\left(\lam\alam
  S \hu \hd+\frac{1}{3}\kap\akap  S^3+\textrm{h.c.}\right)\,,
\label{eq:soft}
\end{eqnarray}
where \lam\ and \kap\ are dimensionless couplings and \alam\ and
\akap\ are trilinear soft parameters. In the CNMSSM-NUHM the soft
masses of the Higgs fields $m_{H_u}$, 
$m_{H_d}$ and $m_S$ are separated from the unified scalar mass
parameter \mzero\ at the grand unification (GUT) scale. These three masses
can be traded at the electroweak (EW) scale for the parameters \tanb\
($\equiv v_u/v_d$, with $v_u$ being the vacuum expectation value (VEV)
of the $u$-type Higgs doublet and $v_d$ that of the $d$-type one),
$\mu_{\rm eff}$ ($\equiv \lam\ s$, with $s$ being the VEV of the
singlet field) and \kap. Similarly \alam$^*$ and \akap$^*$ (with the $^*$
implying that these are defined at the GUT scale) are also disunified from the
 trilinear coupling parameter \azero. The CNMSSM-NUHM thus contains a total of nine continuous
 input parameters, which are given in table~\ref{tab:params} along
 with their ranges scanned for this study. These ranges correspond to 
the `naturalness limit' of the model, where $H_2$ with a mass
consistent with that of the Higgs boson discovered at the 
LHC\cite{Chatrchyan:2012ufa,Aad:2012tfa} can be obtained without
requiring large radiative corrections from the stop sector. 

The tree-level mass-squared of the $A_1$ in the NMSSM is
written in terms of the above parameters (defined at
the SUSY-breaking scale), assuming negligible singlet-doublet mixing, as
\begin{equation}
\label{eq:ma2}
m_{A_1}^2 \simeq \frac{\alam}{2s}  v^2 \lam\sin 2\beta + \kap (2  v^2 \lam
\sin 2\beta - 3s\akap)\,,
\end{equation}
 where $v \equiv \sqrt{v_u^2 + v_d^2} \simeq 174$\gev. At the LHC, the
 $A_1$ can either be produced directly,
preferably in the $gg \rightarrow bbA_1$ channel, owing to the
possibility of a considerably enhanced $b\bar{b}A_1$
coupling\cite{Munir:2013wka} compared to the $ggA_1$ effective
coupling. Alternatively, each of $H_i$, produced in the gluon-fusion
(GF) mode, can also decay into $A_1A_1$ or $A_1Z$ pairs, when kinematically allowed. 
Here we will consider only these indirect production modes.
 
In particular, in case of the decaying SM-like $H_2$, the mass measurement of 
$\sim 125$\gev\ serves as an 
important kinematical handle. Removing this condition (for
$H_1$ and $H_3$) reduces the sensitivity by a factor of $2$ to $3$. 
The $A_1A_1$ pair thus produced 
decays via the $b\bar b b\bar b$ (4$b$), $b\bar b\tau^+\tau^-$ (2$b$2$\tau$) and
$\tau^+\tau^- \tau^+\tau^-$ (4$\tau$) final state combinations. 
In the case of $A_1Z$ production, we only take the $Z\rightarrow \ell^+\ell^-$
decay into account, where $\ell^+\ell^-$ ($2\ell$) stands for $\mu^+\mu^-$ and
$e^+e^-$ combined.

\begin{table}[tbp]
\begin{center}
\begin{tabular}{|c|c|c|c|c|}
\hline
 Parameter & \mzero\,(GeV)  & \mhalf\,(GeV) & \azero\,(GeV) & \mueff\,(GeV)  \\
\hline  
Range  & 200 -- 2000  & 100 -- 1000 & $-3000$ -- 0 & 100 -- 200 \\
\hline
\hline
\tanb & \lam  & \kap  & \alam$^*$\,(GeV)  & \akap$^*$\,(GeV) \\
\hline 
1 -- 6 & 0.4 -- 0.7  & 0.01 -- 0.7 & $-500$ -- 500  & $-500$ -- 500 \\
\hline
\end{tabular}
\caption{The CNMSSM-NUHM input parameters and their scanned ranges.}
\label{tab:params}
\end{center}
\end{table}

\section{\label{method} Parameter scans and event analysis}

We scanned the NMSSM parameter space, given in table\,\ref{tab:params}, to search for
regions yielding $m_{A_1} \lesssim 150$\gev\ and the mass of $H_2$, $m_{H_2}$, around
125\gev. We used the publicly available package 
NMSSMTools-v4.2.1\cite{NMSSMTools} for computation of the SUSY
mass spectrum and branching ratios (BR) of the Higgs bosons for each
model point.  In our scans we imposed the constraints from 
$b$-physics, based on \cite{Beringer:1900zz}, and from Dark Matter relic density
measurement\cite{Ade:2013zuv}, as
\begin{itemize}
\item $\brbsmumu = (3.2~(\pm 10\%~{\rm theoetical~error})\pm1.35) \times 10^{-9}$,
\item $\brbutaunu = (1.66\pm 0.66 \pm 0.38) \times 10^{-4}$,
\item $\brbxsgamma = (3.43\pm 0.22 \pm0.21) \times 10^{-4}$,
\item $\Omega_\chi h^2 < 0.131~(0.119+10\%~{\rm theoetical~error})$.
\end{itemize}
Exclusion limits from the LEP and LHC Higgs boson searches were also tested
against using the HiggsBounds-v4.1.3\cite{Bechtle:2013wla} package.
Finally, from NMSSMTools we obtained the signal rates of $H_2$, defined
for a given decay channel $X$ as
\begin{equation}
\label{eq:rggh}
R_X \equiv  \frac{\sigma(gg\rightarrow H_2)\times {\rm BR}(H_i\rightarrow
  X)}{\sigma(gg\rightarrow h_{\rm SM})\times {\rm BR}(h_{\rm SM} \rightarrow X)}\,,
\end{equation}
where ${h_{\rm SM}}$ is the SM Higgs boson with the same mass as
$H_2$. We then required $R_X$ for $X=\gamma\gamma,\,ZZ$ to lie within 
the measured $\pm 1\sigma$ ranges of the corresponding experimental
quantities $\mu_X$ by the CMS collaboration\cite{CMS-PAS-HIG-14-009}. 
These ranges read
\begin{equation}
\label{eq:CMS-mu}
\mu_{\gamma \gamma} = 1.13 \pm 0.24~~{\rm and}~~\mu_{ZZ} = 1.0 \pm 0.29\,.
\end{equation}

Following the scans, we carried out a dedicated
signal-to-background analysis based on Monte Carlo event
generation for proton-proton collisions at 14\,TeV centre-of-mass
energy at the LHC, for each process of interest. 
Using the program SuSHi-v1.1.1\cite{Harlander:2012pb}, we
first calculated the GF production cross section
of an SM Higgs boson with the same mass as as that of a $H_i$ which 
is expected to decay into $A_1A_1$ or $A_1Z$ for a given SUSY point. 
This cross section was then rescaled using the $ggH_i$
reduced coupling in the NMSSM, and multiplied by the relevant BRs of
the $H_i$, all of which are obtained from NMSSMTools. 
The backgrounds, which include the $pp\to 4b$,
$pp\to 2b2\tau$, $pp\to 4\tau$,
$pp\to Z 2b$ and $pp\to Z2\tau$ processes, were computed
with MadGraph 5\cite{Alwall:2011uj}. Both the signal and the
background for each process were hadronised and
fragmented using Pythia 8.180\cite{Sjostrand:2007gs} interfaced with
 FastJet-v3.0.6\cite{Cacciari:2011ma} for jet clustering. The parton-level acceptance cuts
used are
\begin{itemize}
\item $|\eta|<$ 2.5 for all final state objects,
\item $p_T>15$\gev\ far all final state objects,
\item $\Delta R \equiv \sqrt{(\Delta\eta)^2+(\Delta\phi)^2}>0.2$ for all $b$-quark pairs,
\item $\Delta R >0.4$ for all other pairs of final state objects,
\end{itemize}
where $p_T$, $\eta$, $\phi$ are the transverse momentum,
pseudorapidity and azimuthal angle, respectively. 

Our use of the jet substructure method\cite{Butterworth:2008iy}
 implied that we had three possible signatures for a decaying $A_1$: one
 fat jet, two single $b$-jets and two $\tau$-jets. The fat jet
 analysis, which assumes boosted $b$-quarks, allows one to obtain 
much higher sensitivities, particularly
 for large masses of the decaying Higgs bosons. We then calculated
the expected cross sections for the signal processes which
yield $S/\sqrt{B}>5$ for three benchmark accumulated luminosities at
the LHC, $\mathcal{L}=30$/fb,\,300/fb and 3000/fb, in various final
state combinations, as functions of $m_{A_1}$. 

\section{\label{results} Results}

For the figures shown in this section, we first make two assertions: 
1) all the points shown satisfy the constraints mentioned
earlier and yield $122\gev\ < m_{H_2} <129\gev$, and 2) the sensitivity curve(s) shown corresponds to the best
final state combination(s) for probing the given process. \\

\begin{figure}[tbp]
\centering
\subfloat[]{%
\label{fig:-a}%
\includegraphics*[width=7cm]{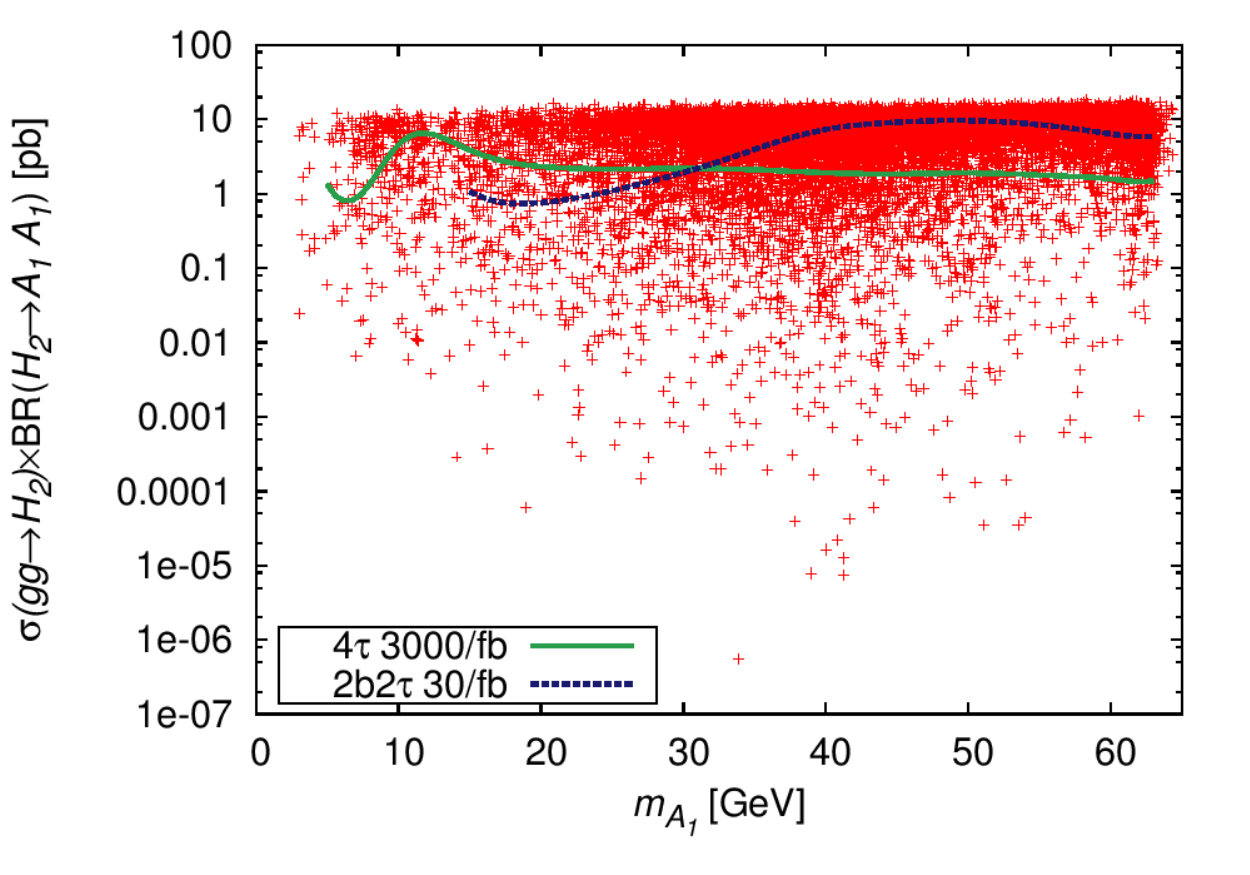}
}%
\hspace{0.5cm}%
\subfloat[]{%
\label{fig:-b}%
\includegraphics*[width=7cm]{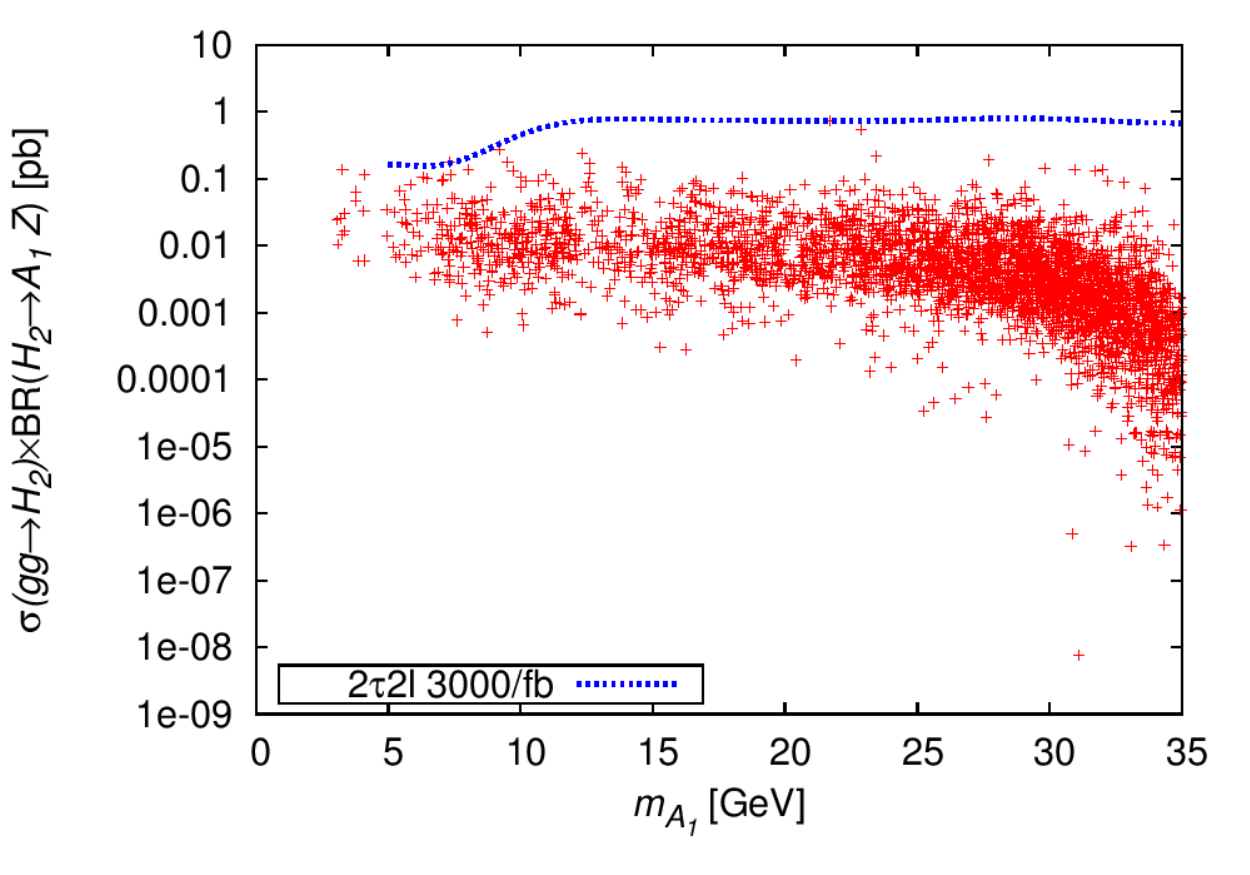}
}
\caption{Cross sections for (a) the $gg \to H_2\to  A_1A_1$ process and (b)
 the $gg \to H_2\to A_1Z$ process.}
\label{fig:H2toA1}
\end{figure}

\noindent \underline{\it Production via $H_2 \rightarrow A_1A_1/Z$}: We begin with the
decays of the SM-like $H_2$, since reconstructing its correct mass
improves the kinematics, as noted earlier. 
In figure~\ref{fig:H2toA1}(a) we show the prospects for the $H_2 \to
A_1A_1$ channel at the LHC. Also shown are the sensitivity curves 
for the $2b2\tau$ final state at $\mathcal{L}=30$/fb and for the
$4\tau$ final state at $\mathcal{L}=3000$/fb.
We see that a large part of the NMSSM parameter space can be probed 
via $H_2\rightarrow A_1 A_1$ decays at the LHC, at $\mathcal{L}$ as
low as 30/fb. Note
that the Higgs boson signal rate constraints from CMS restrict the BR($H_2 \to
A_1A_1$) to less than 50\%. In figure~\ref{fig:H2toA1}(b) we
see that the $H_2\to A_1Z$ decay shows no promise even at $\mathcal{L}=3000$/fb.  \\

\begin{figure}[tbp]
\centering
\subfloat[]{%
\label{fig:-a}%
\includegraphics*[width=7cm]{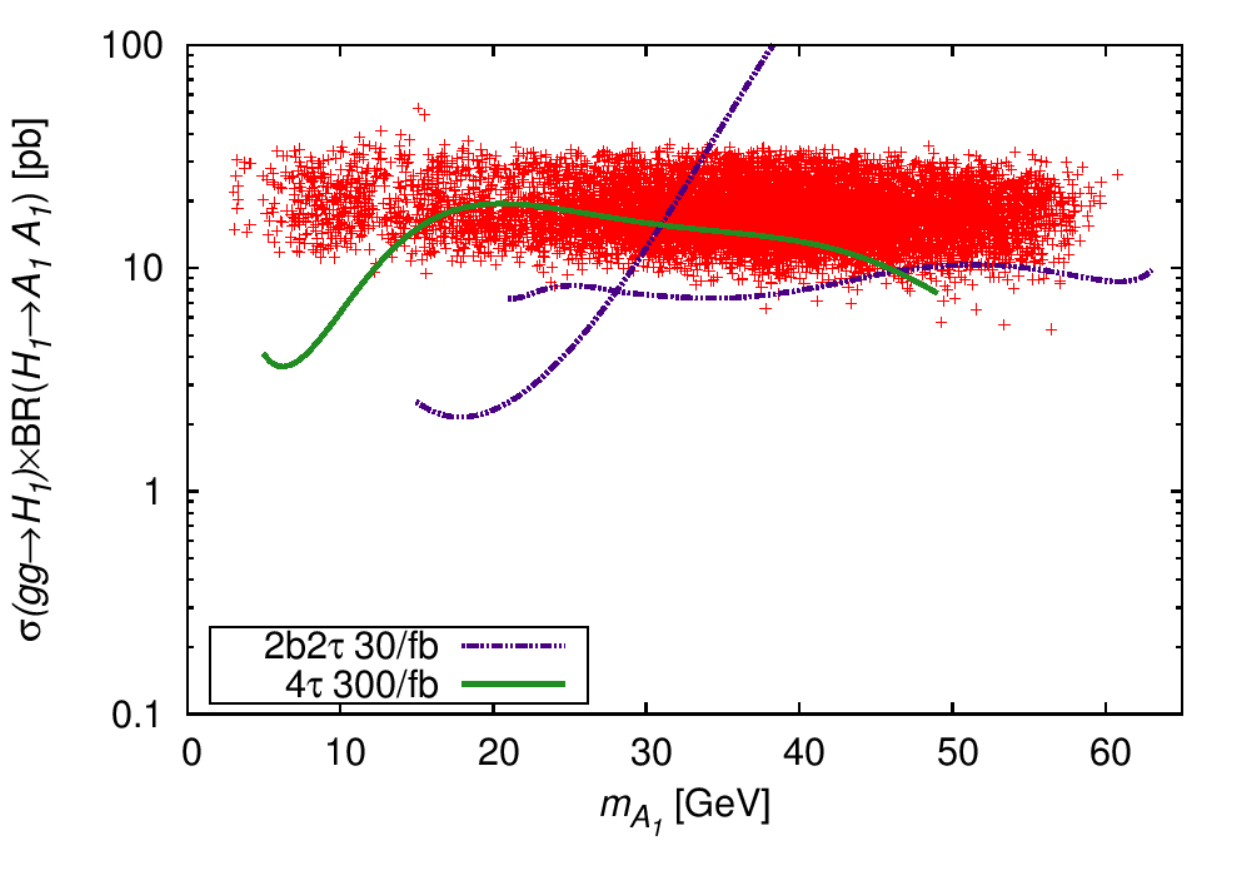}
}%
\hspace{0.5cm}%
\subfloat[]{%
\label{fig:-b}%
\includegraphics*[width=7cm]{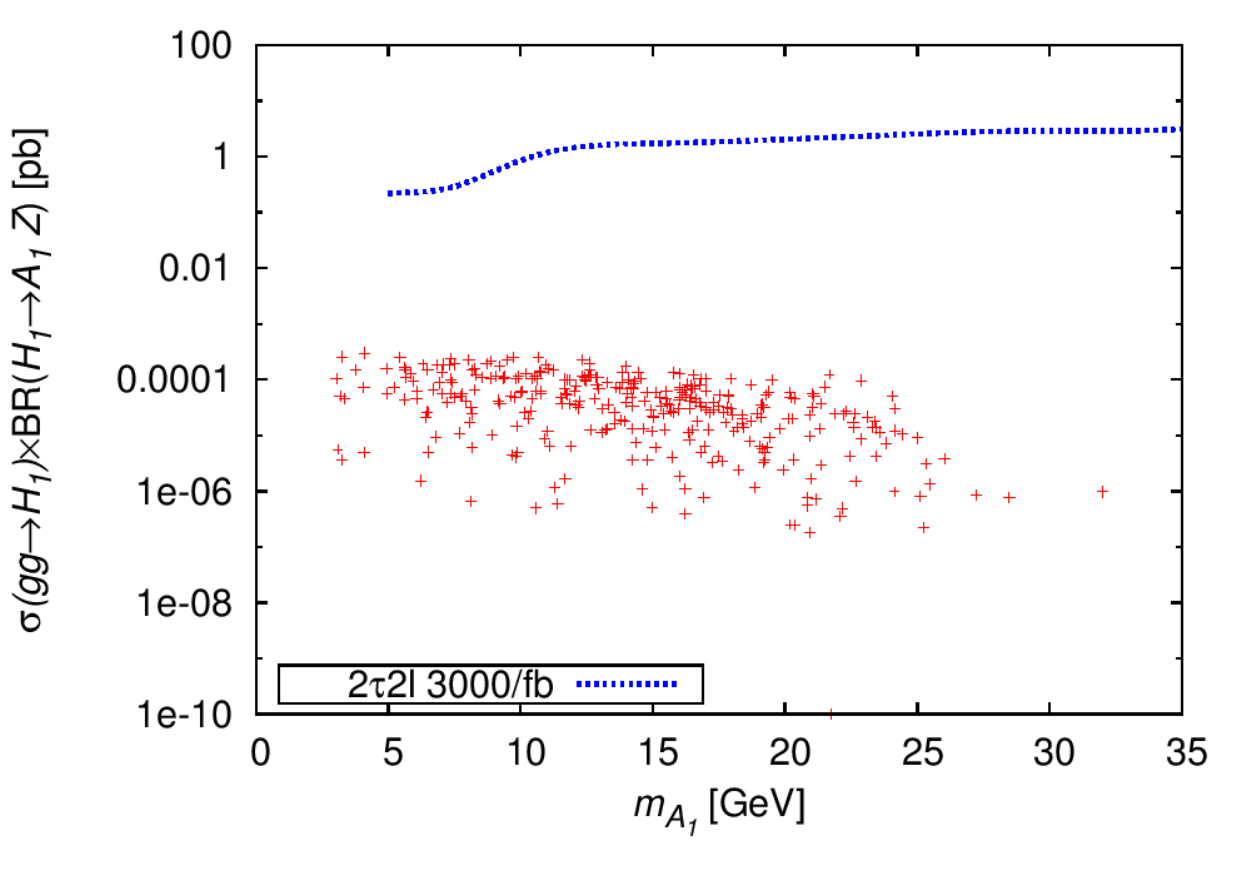}
}
\caption{Cross sections for (a) the $gg \to H_1 \to  A_1A_1$ process and (b)
 the $gg \to H_1 \to A_1Z$ process.}
\label{fig:H1toA1}
\end{figure}

\noindent \underline{\it Production via $H_{1,3} \rightarrow
  A_1A_1/Z$}: The case of the $H_1 \to
A_1A_1$ decay, for a singlet-like $H_1$, is illustrated in
figure~\ref{fig:H1toA1}(a). One sees that almost
all the points with $m_A \gtrsim 12$\gev\
are potentially discoverable in the $2b2\tau$ final state at
$\mathcal{L}=30$/fb. Two separate sensitivity curves corresponding to
this final state indicate that for low $A_1$ masses the fat jet
analysis has been employed, which results in a better reach. 
Even lighter $A_1$ could also be visible in the
$4\tau$ final state with $\mathcal{L}=300$/fb.
Figure~\ref{fig:H1toA1}(b) shows poor prospects for
the discovery of $A_1$ via the $H_1\to A_1 Z$ channel also. 

\begin{figure}[tbp]
\centering
\subfloat[]{%
\label{fig:-a}%
\includegraphics*[width=7cm]{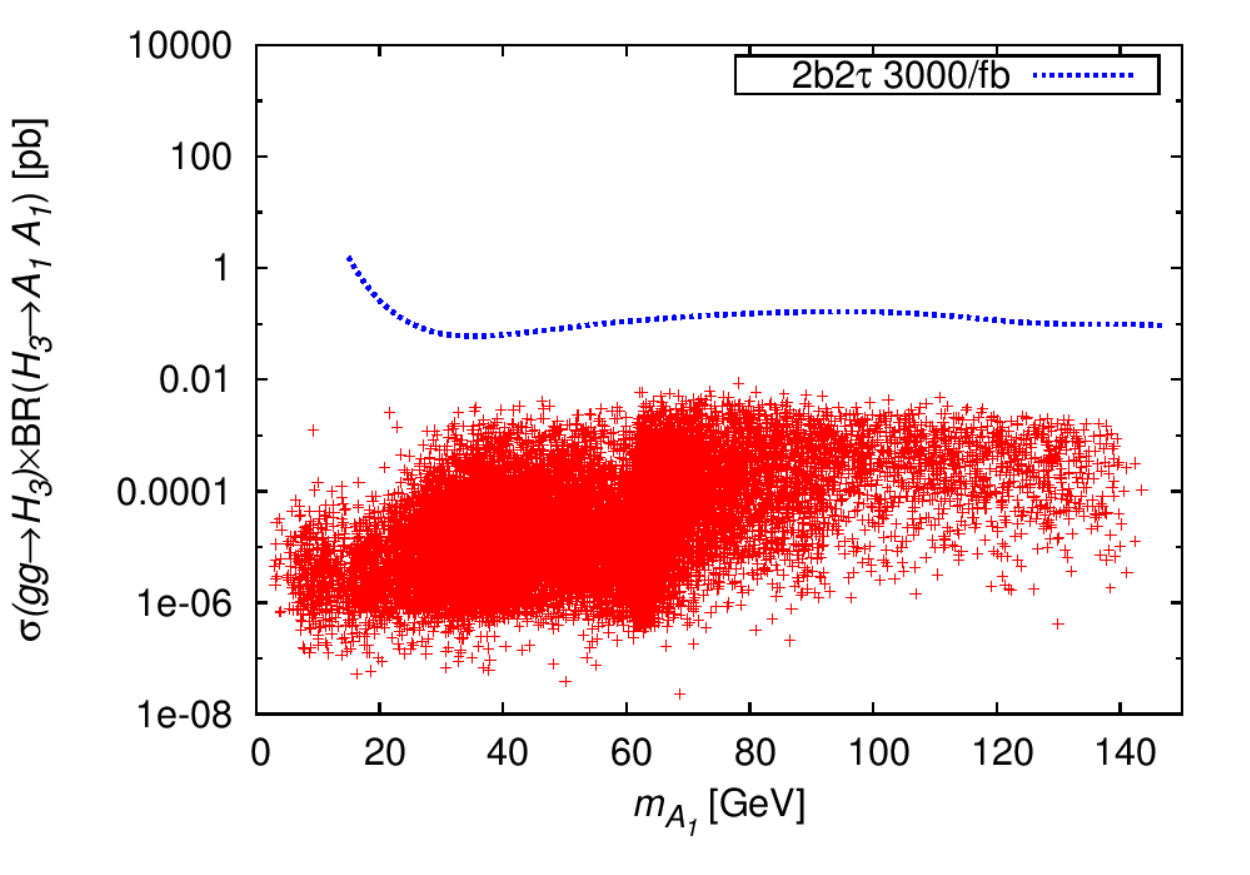}
}%
\hspace{0.5cm}%
\subfloat[]{%
\label{fig:-b}%
\includegraphics*[width=7cm]{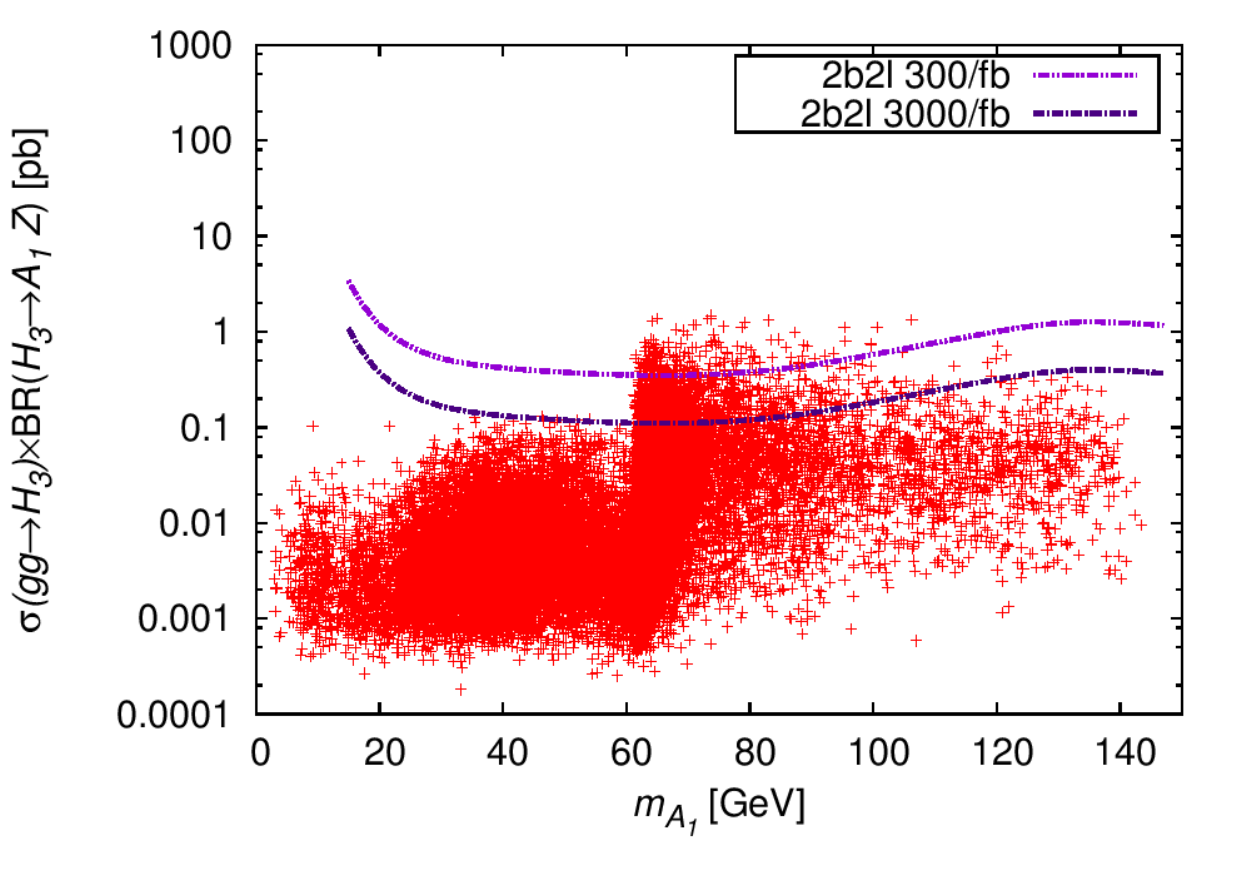}
}
\caption{Cross sections for (a) the $gg \to H_3 \to  A_1A_1$ process and (b)
 the $gg \to H_3 \to A_1Z$ process.}
\label{fig:H3toA1}
\end{figure}

In figure~\ref{fig:H3toA1}(a) we see that the $H_3\to A_1A_1$ channel
will be inaccessible at the LHC due to the fact that for such high
masses of $H_3$ ($\gtrsim 400$\,GeV) the
production cross section gets diminished. Moreover, other decay channels of
$H_3$ dominate over this channel. The sensitivity curve in the figure
corresponds to the $2b2\tau$ final state for $\mathcal{L}=3000$/fb. 
Conversely, as shown in figure~\ref{fig:H3toA1}(b), in the $H_3\to
A_1Z$ channel a number of points lie above the $2b2\ell$
sensitivity curve for $\mathcal{L}=300$/fb. The discoverability of an
$A_1$ in this channel results from the use
of the fat jet analysis as well as from a sizeable $H_3A_1Z$
coupling, owing to a significant doublet component in $A_1$.

In summary, the decays of the two lightest scalar Higgs bosons 
carry the potential to reveal an $A_1$ with mass 
$\lesssim 60$\gev\ for an integrated luminosity of 30/fb at the LHC.
When the $A_1$ is heavier than $\sim 60$\gev, while its
pair production also becomes inaccessible, the $gg\to H_3\to A_1Z$ channel
takes over as the most promising one. This channel is, therefore, of great
importance and warrants dedicated probes in future analyses at the LHC. 

\acknowledgments{This work has been funded in part by the Welcome Programme
of the Foundation for Polish Science. S.~Moretti is supported in part through the NExT
Institute. S.~Munir is supported in part by the Swedish Research
Council under contracts 2007-4071 and 621-2011-5107. The use of the
CIS computer cluster at NCBJ is gratefully acknowledged.}

\end{document}